\documentstyle[emulateapj,apjfonts]{article}
\newcommand{\preprint}[1]{#1}

\lefthead{Becker, White, et al.}
\righthead{Radio-Selected BAL Quasars}

\begin{document}

\title{Properties of Radio-Selected Broad Absorption-Line Quasars
from the FIRST Bright Quasar Survey} 

\author{
Robert~H.~Becker\altaffilmark{1,2,3},
Richard~L.~White\altaffilmark{3,4},
Michael~D.~Gregg\altaffilmark{1,2,3},
Michael~S.~Brotherton\altaffilmark{2,3},
Sally~A.~Laurent-Muehleisen\altaffilmark{1,2,3},
Nahum~Arav\altaffilmark{2}}
\authoremail{bob@igpp.llnl.gov}

\altaffiltext{1}{Physics Dept., University of California, Davis, CA
95616, bob@igpp.llnl.gov}
\altaffiltext{2}{IGPP/Lawrence Livermore National Laboratory}
\altaffiltext{3}{Visiting Astronomer, Kitt Peak National
Observatory, National Optical Astronomy Observatory}
\altaffiltext{4}{Space Telescope Science Institute}

\begin{abstract}

In a spectroscopic follow-up to the VLA FIRST survey, the FIRST Bright
Quasar Survey (FBQS) has found 29 radio-selected broad absorption line
(BAL) quasars.  This sample provides the first opportunity to study
the properties of radio-selected BAL quasars.  Contrary to most
previous studies, we establish that a significant population of
radio-loud BAL quasars exists.  Radio-selected BAL quasars display
compact radio morphologies and possess both steep and flat radio
spectra.  Quasars with low-ionization BALs have a color
distribution redder than that of the FBQS sample as a whole.  The
frequency of BAL quasars in the FBQS is significantly greater, perhaps by as much as
factor of two, than that inferred from optically selected samples.
The frequency of BAL quasars appears to have a complex dependence on
radio-loudness.  The properties of this sample appear inconsistent
with simple unified models in which BAL quasars constitute
a subset of quasars seen edge-on.

\end{abstract}


\section{Introduction}
\label{sectionintro}

Until recently, a search of the astronomical literature would have
revealed that broad absorption lines (BAL) are seen in approximately
10\% of optically selected quasars (\cite{foltz90}; \cite{weymann91})
and in exactly 0\% of radio-loud quasars (\cite{stocke92}).
This dichotomy has puzzled astronomers for years. The BAL quasars can
be divided into two classes, high-ionization and low-ionization, which
are primarily defined by the presence of broad absorption by C~IV
$\lambda1549$ and Mg~II $\lambda2800$, respectively. (Note that all
low-ionization BAL quasars also show high ionization absorption). The
high-ionization BAL (HiBAL) quasars are more common, including 10\% of
all optically selected quasars, while the rarer low-ionization BAL
(LoBAL) quasars make up only 1\% of optically selected quasars. 

Prior to the FIRST survey there was only a single example of a LoBAL quasar whose
spectrum shows strong absorption by metastable excited states of Fe~II
(Q 0059$-$2735, \cite{hazard87}).  Becker et al.\ (1997) reported the
discovery of two more objects resembling Q0059$-$2735 (FIRST
J084044.5+363328 and J155633.8+351758), the second of which is
radio-loud. We will refer to these as FeLoBAL quasars.  Both of the
new unusual quasars were found by making optical identifications of
radio sources from the VLA FIRST survey (Faint Images of the Radio Sky
at Twenty-cm, \cite{becker95}; \cite{white97}).  Subsequently,
Brotherton et al.\ (1998) identified five more radio-loud BAL quasars
(two HiBAL and three LoBAL quasars) from a complete sample of
radio-selected ultraviolet excess quasars, firmly establishing the
existence of radio-loud BAL quasars.  Lastly, Wills, Brandt, and Laor
(1999) have recently suggested that the radio-loud quasar PKS 1004+13
is also a BAL quasar.

For the past five years we have been developing several new
radio-selected samples of quasars based on the VLA FIRST survey. The
most extensive of these is the FIRST Bright Quasar Survey or FBQS
(Gregg et al.\ 1996, hereafter \cite{gregg96}; White et al.\ 2000,
hereafter \cite{white99}).  The goal of the FBQS is to identify all
quasars in the FIRST survey brighter than 17.8 on the POSS-I $E$ (red)
plate. In the initial 2700 square degrees of the FIRST survey, we
defined a sample of 1238 quasar candidates based on positional
coincidence between a FIRST source and a POSS-I stellar object (see
\cite{gregg96} and \cite{white99} for a detailed discussion of the
candidate selection criteria).  Spectra have been collected for 90\%
of these candidates, 636 of which have been identified as
quasars.  Among these are 29 which display BAL characteristics.  We
present the optical spectra and radio spectral indices of these BAL
quasars, comparing their radio and optical properties to previous
samples of optically selected BAL quasars.  We discuss the selection
biases inherent in the survey results and discuss why our sample
differs from those based on optically selected samples.

\section{Identification and Classification of BAL Quasars}

The FBQS BAL quasars are defined to be any quasar which shows
significant broad absorption blueward of either Mg~II $\lambda2800$ or
C~IV $\lambda1549$.  We have chosen not to employ any strict
definition of a BAL, such as the ``BALnicity'' index of Weymann et
al.\ (1991), which requires continuous absorption of at least 10\% in
depth spanning more than 2000~km~s$^{-1}$, discounting absorption
closer than 3000~km~s$^{-1}$ blueward of the emission peak.  Weymann's
highly conservative definition has the advantage of unambiguously
distinguishing between associated absorbers and ``classical BALs'' but
could unnecessarily exclude several potentially very interesting
members of the class.  We advocate classification of absorption
systems based on their physical characteristics such as variability
and partial coverage (c.f.\ \cite{barlow97}) and so we choose not to
exclude any likely BAL quasars.


Even though we did not use ``BALnicity'' to define our sample, we have
calculated the BALnicity index for each of our BALs.  The values are
given in Table~1. In all, seven of our BALs fail the BALnicity test,
i.e., they have zero BALnicity. Three of these are LoBALs for which
by necessity the BALnicity was calculated from the Mg~II absorption line, a line
for which the test was never meant to be applied (Weymann et al 1991).
For example, Voit et al (1993) found that Mg~II absorption troughs are
usually narrower than the C~IV absorptions troughs in LoBALs. Two of these LoBALs are in
fact FeLoBALs and the correctness of their inclusion is almost beyond
question in so far as the conditions necessary for absorption by excited states
of Fe strongly indicate an intrinsic system local to the active 
nucleus (FIRST J084044.5+363328 and FIRST 121442.3+280329).  The
inclusion of the third (FIRST J112220.5+312441) is problematical and
will only be resolved with an observation of C~IV.  Inclusion of the
four HiBALs which fail the test can be justified as follows.  Two of
them (FIRST J095707.4+235625 and J141334.4+421202) have nearly black
C~IV absorption spanning 4000 km/sec which is very unlikely to break
up into a blend of narrow lines.  FIRST~J115023.6+281908 has three C~IV
absorption systems with velocities up to 11700 km/sec.  Even though
none of the absorbers individually is 2000 km/sec broad, the three
taken together are very suggestive of being an intrinsic BAL outflow.  Lastly, the Si~IV
absorption lines in FIRST~J160354.2+300209 show clear evidence of
partial covering which is normally taken to be a property of BALs
(Arav et al.\ 1999).
%

The typical wavelength coverage of the FBQS spectra is 3800 to
8000~\AA.  For quasars with $0.5 < z < 1.7$, the MgII~2800 feature is
shifted into the observed range, permitting identification of LoBALs.
Somewhat higher redshift LoBALs can be identified through broad
absorption by Al~III $\lambda1860$ as in the case of
FIRST~J105427.1+253600.  HiBAL quasars can be identified only for $z
\gtrsim 1.4$ which brings CIV $\lambda$1549 well into the observed
spectral range.  (Since the observed wavelength coverage is not
uniform for all the FBQS spectra, the redshift range over which C~IV
is observable differs from quasar to quasar.) Some of our HiBAL
quasars may actually be unrecognized LoBAL or FeLoBAL quasars, since
LoBALs also exhibit broad CIV $\lambda$1549 and other high-ionization
species.

Table~1, partially excerpted from Table~2 in FBQS2, lists FIRST catalog RA and
Dec (J2000), recalibrated and extinction-corrected $E$ and $O$
magnitudes, red extinction corrections $A(E)$,
FIRST peak and integrated radio flux
densities, the computed BALnicity index (as defined in \cite{weymann91}), the maximum outflow
velocity in the absorption lines,
and redshifts for the 29 BAL quasars identified to date in
the FBQS.  Also in Table~1 are the radio luminosity $L_R$ at a rest
frequency of 5-GHz (calculated using the observed radio spectral indices
from Table~2 and hence different from the values given in FBQS2),
the absolute $B$ magnitude $M_B$, and the radio
loudness, $R^*$, the ratio of the 5~GHz radio flux density to the
2500~\AA\ optical flux in the quasar rest frame (using $\alpha_{radio}$
from Table 2 and assuming
$\alpha_{opt} = -1$, \cite{stocke92}).  We use the (APS-calibrated) APM
$O$ magnitude (White et al.\ 2000) as a direct estimate of $B$, and we
do not correct the optical magnitude for the emission-line
contribution.  The cosmological parameters $H_{\rm
o}=50\,\hbox{km}\,\hbox{s}^{-1}\hbox{Mpc}^{-1}$, $\Omega=1$, and
$\Lambda=0$ are adopted.  In the last column, we give the type of
BAL.  Our spectra for FIRST~J112220.5+312441 and J115023.6+281908, while suggestive that
these objects are BAL QSOs, are not definitive; this uncertainty is
indicated by question marks next to the type in Table~1.
Figures~\ref{figlospectra}
and~\ref{fighispectra} show the spectra of the BAL quasars, plotted in
the rest frame to facilitate the recognition of the sometimes complex
absorption features.

\placetable{table1}


\preprint{
\begingroup
\tabcolsep=3pt
\footnotesize
\begin{deluxetable}{rrccccrrcrrccrl}
\tablewidth{0pt}
\tablecaption{FIRST Broad Absorption Line Quasars}
\tablenum{1}
\tablehead{
\colhead{RA}&\colhead{Dec}&\colhead{E}&\colhead{O}&\colhead{O-E}&\colhead{A(E)}&\colhead{$S_p$}&\colhead{$S_i$}&\colhead{BALnicity}&\colhead{$V_{max}$}&\colhead{$M_B$}&\colhead{$\log L_R$}&\colhead{$\log R^*$}&\colhead{z}&\colhead{Notes}\\
\colhead{(1)}&\colhead{(2)}&\colhead{(3)}&\colhead{(4)}&\colhead{(5)}&\colhead{(6)}&\colhead{(7)}&\colhead{(8)}&\colhead{(9)}&\colhead{(10)}&\colhead{(11)}&\colhead{(12)}&\colhead{(13)}&\colhead{(14)}&\colhead{(15)}
}
\startdata
07 24 18.492&+41 59 14.40&  17.65&  18.71&  1.05&0.24&       7.89&      7.90& \phn1300    & 18300 &-26.6& 32.7&  1.60& 1.552&LoBAL\nl
07 28 31.661&+40 26 15.85&  15.13&  15.26&  0.14&0.13&      16.96&     16.79& \phn\phn200 & 20600 &-28.0& 32.0&  0.37& 0.656&LoBAL\nl
08 09 01.332&+27 53 41.67&  17.17&  17.58&  0.40&0.10&       1.17&      1.67& \phn7000       & 27400 &-27.7& 31.9&  0.40& 1.511&HiBAL\nl
08 40 44.457&+36 33 28.41&  16.20&  17.53&  1.34&0.10&       1.63&      1.00& \phn\phn\phn\phn0 &  4900 &-27.2& 31.8&  0.45& 1.230&FeLoBAL\nl
09 10 44.902&+26 12 53.71&  17.78&  19.26&  1.48&0.09&       7.84&      7.46& \phn\phn320 &  6100 &-27.6& 33.1&  1.64& 2.920&HiBAL\nl
09 13 28.260&+39 44 44.17&  17.16&  17.99&  0.83&0.05&       2.06&      2.09& 10700       & 27000 &-27.4& 32.0&  0.63& 1.580&HiBAL\nl
09 34 03.978&+31 53 31.47&  17.43&  17.83&  0.39&0.05&       4.68&      4.41& \phn1200    & 27000 &-28.5& 32.8&  0.88& 2.419&HiBAL\nl
09 46 02.299&+27 44 07.04&  16.89&  17.65&  0.76&0.06&       3.54&      3.63& \phn\phn260 &  9900 &-27.9& 32.4&  0.74& 1.748&HiBAL\nl
09 57 07.367&+23 56 25.32&  17.63&  18.47&  0.84&0.09&     136.10&    140.49& \phn\phn\phn\phn0 &  5200 &-27.4& 34.1&  2.63& 1.995&HiBAL\nl
10 31 10.647&+39 53 22.81&  17.73&  18.66&  0.92&0.03&       2.45&      2.03& \phn\phn\phn20 &  5900 &-25.8& 31.9&  1.11& 1.082&LoBAL\nl
10 44 59.591&+36 56 05.39&  16.51&  17.23&  0.72&0.04&      14.61&     15.00& \phn\phn400 &  6600 &-26.2& 32.2&  1.29& 0.701&FeLoBAL\nl
10 54 27.150&+25 36 00.33&  16.93&  18.38&  1.45&0.07&       2.99&      3.02& \phn1200\tablenotemark{a}  & 25000 &-28.0& 32.6&  0.91& 2.400&LoBAL\nl
11 22 20.462&+31 24 41.19&  17.08&  18.19&  1.11&0.04&      12.64&     12.87& \phn\phn\phn\phn0 &  4300 &-26.9& 32.8&  1.52& 1.448&LoBAL?\tablenotemark{b}\nl
11 50 23.570&+28 19 07.50&  16.46&  18.00&  1.54&0.06&      13.96&     14.22& \phn\phn\phn\phn0 & 11700 &-29.0& 33.5&  1.43& 3.124&HiBAL?\tablenotemark{c}\nl
12 00 51.501&+35 08 31.41&  15.21&  16.45&  1.24&0.05&       2.03&      1.46& \phn4600    & 13500 &-29.1& 32.1& -0.02& 1.700&HiBAL\nl
12 14 42.303&+28 03 29.01&  15.87&  17.03&  1.16&0.06&       2.61&      2.90& \phn\phn\phn\phn0 &  3500 &-26.3& 31.4&  0.39& 0.698&FeLoBAL\nl
13 04 25.543&+42 10 09.66&  16.33&  17.08&  0.75&0.04&       1.52&      0.99& \phn2900    & 27000 &-28.7& 32.2&  0.23& 1.916&HiBAL\nl
13 12 13.560&+23 19 58.51&  16.95&  17.84&  0.89&0.03&      43.27&     44.12& \phn1400    & 25000 &-27.4& 33.3&  1.88& 1.508&HiBAL\nl
13 24 22.536&+24 52 22.25&  17.40&  18.87&  1.47&0.04&       4.89&      4.41& \phn1300\tablenotemark{a}  &  6900 &-27.4& 32.7&  1.31& 2.357&LoBAL\nl
14 08 00.454&+34 51 25.11&  17.03&  18.44&  1.41&0.04&       2.91&      2.87& \phn\phn260    &  9600 &-26.3& 31.9&  0.99& 1.215&LoBAL\nl
14 08 06.207&+30 54 48.67&  17.30&  19.00&  1.69&0.03&       3.34&      3.21& \phn4800       & 22000 &-24.8& 31.7&  1.28& 0.842&LoBAL\nl
14 13 34.404&+42 12 01.76&  17.60&  18.40&  0.80&0.02&      17.79&     18.74& \phn\phn\phn\phn0 &  3800 &-28.3& 33.5&  1.68& 2.810&HiBAL\nl
14 20 13.072&+25 34 03.71&  16.78&  18.24&  1.47&0.05&       1.27&      1.17& \phn4500       & 26000 &-27.9& 32.1&  0.46& 2.200&HiBAL\nl
14 27 03.637&+27 09 40.29&  17.68&  19.05&  1.37&0.05&       2.58&      2.98& \phn\phn\phn30 &  5900 &-25.6& 31.9&  1.23& 1.170&FeLoBAL\nl
15 23 14.434&+37 59 28.71&  17.37&  18.41&  1.04&0.04&       1.67&      1.83& \phn3700    & 17000 &-26.5& 31.8&  0.76& 1.344&HiBAL\nl
15 23 50.435&+39 14 04.83&  15.93&  16.93&  1.00&0.05&       3.75&      4.07& \phn3700    & 19000 &-26.3& 31.6&  0.66& 0.657&LoBAL\nl
16 03 54.159&+30 02 08.88&  17.36&  18.02&  0.65&0.10&      53.69&     54.18& \phn\phn\phn\phn0 &  3200 &-27.9& 33.7&  2.04& 2.026&HiBAL\nl
16 41 52.295&+30 58 51.79&  17.57&  18.70&  1.14&0.07&       2.14&      2.66& 10600       & 22000 &-27.2& 32.4&  1.09& 2.000&LoBAL\nl
16 55 43.235&+39 45 19.91&  17.74&  18.11&  0.37&0.05&      10.15&     10.16& \phn4500    & 22000 &-27.5& 32.9&  1.41& 1.747&HiBAL\nl
\tablenotetext{a}{The BALnicity was computed from C~IV for these LoBAL quasars. Other LoBAL quasars have the BALnicity
computed from Mg~II, and all HiBAL quasars have BALnicities from C~IV.}
\tablenotetext{b}{Absorption line depth barely exceeds 10\% limit for BALnicity criterion.}
\tablenotetext{c}{Identification as BAL quasar is less secure; see text
for discussion.}
\enddata
\end{deluxetable}
\endgroup

}

\placefigure{figlospectra}

\placefigure{fighispectra}

\preprint{
\begin{figure*}
\epsfxsize=\textwidth \epsfbox{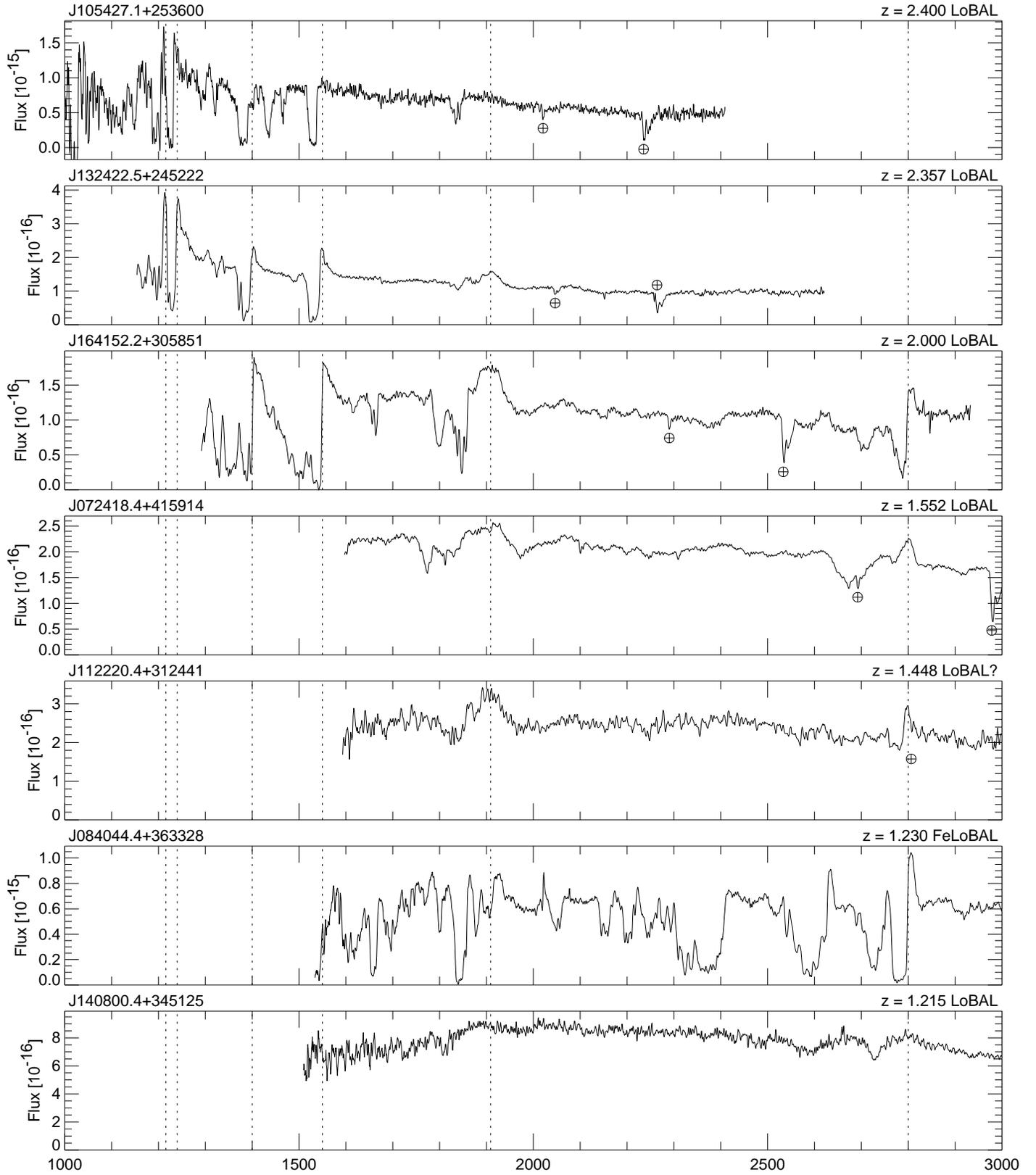}
\caption{ Spectra of low ionization broad absorption line quasars from
the FIRST Bright Quasar Survey, sorted by decreasing redshift, plotted
against rest wavelength.  The dotted lines show expected positions of
prominent emission lines: Ly$\alpha$~1216, N~V~1240, Si~IV~1400,
C~IV~1549, C~III]~1909, and Mg~II~2800.  The positions of the
atmospheric A and B band absorption at observed wavelengths of
$\sim6880$~\AA\ and 7620~\AA\ are marked.
\label{figlospectra}}
\end{figure*}

\begin{figure*}[p]
\figurenum{1}
\epsfxsize=\textwidth \epsfbox{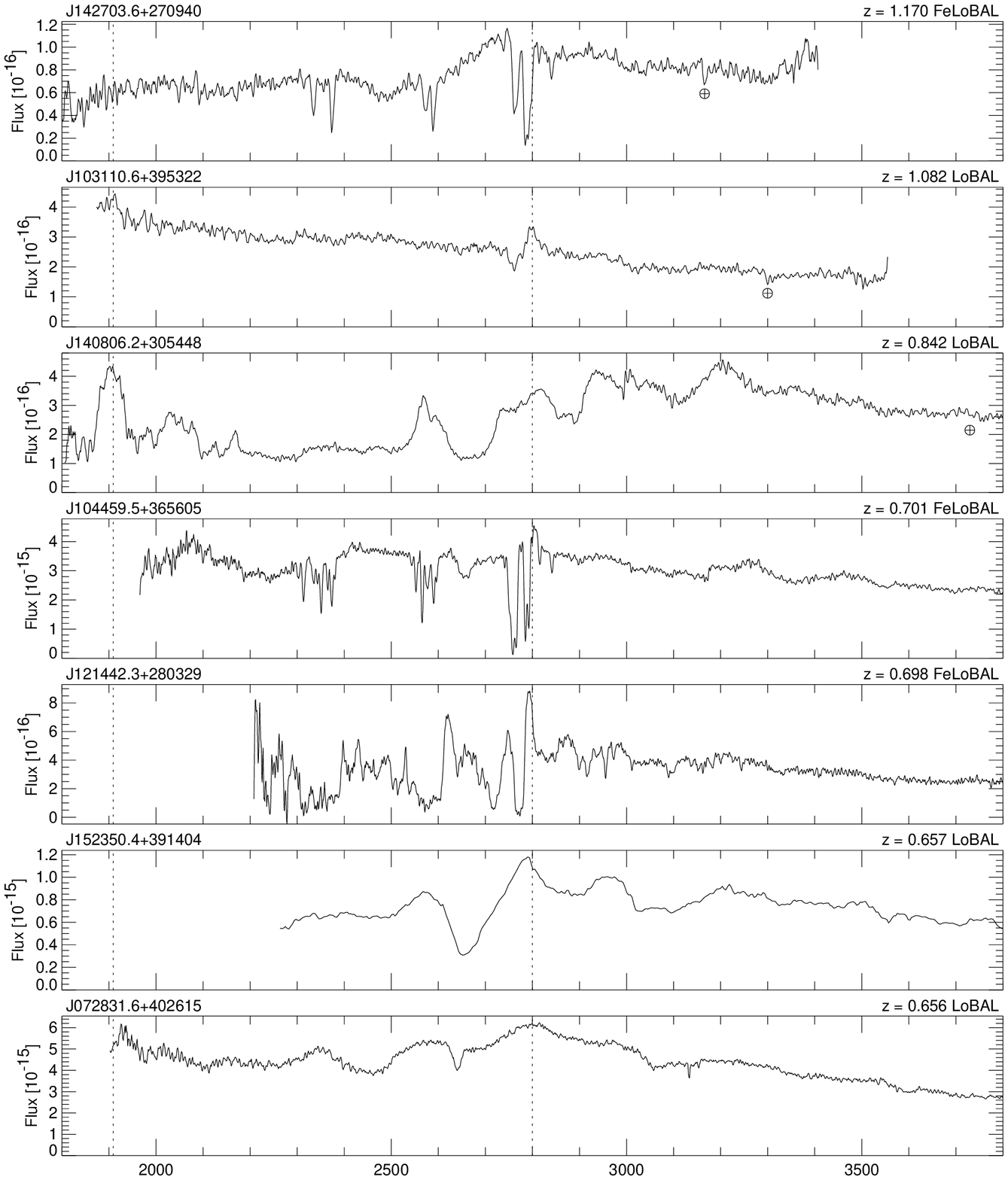}
\caption{{\it Continued.}  Spectra of FBQS low-ionization BAL quasars.
Note that the wavelength range differs from previous panel.}
\end{figure*}

\begin{figure*}[p]
\epsfxsize=\textwidth \epsfbox{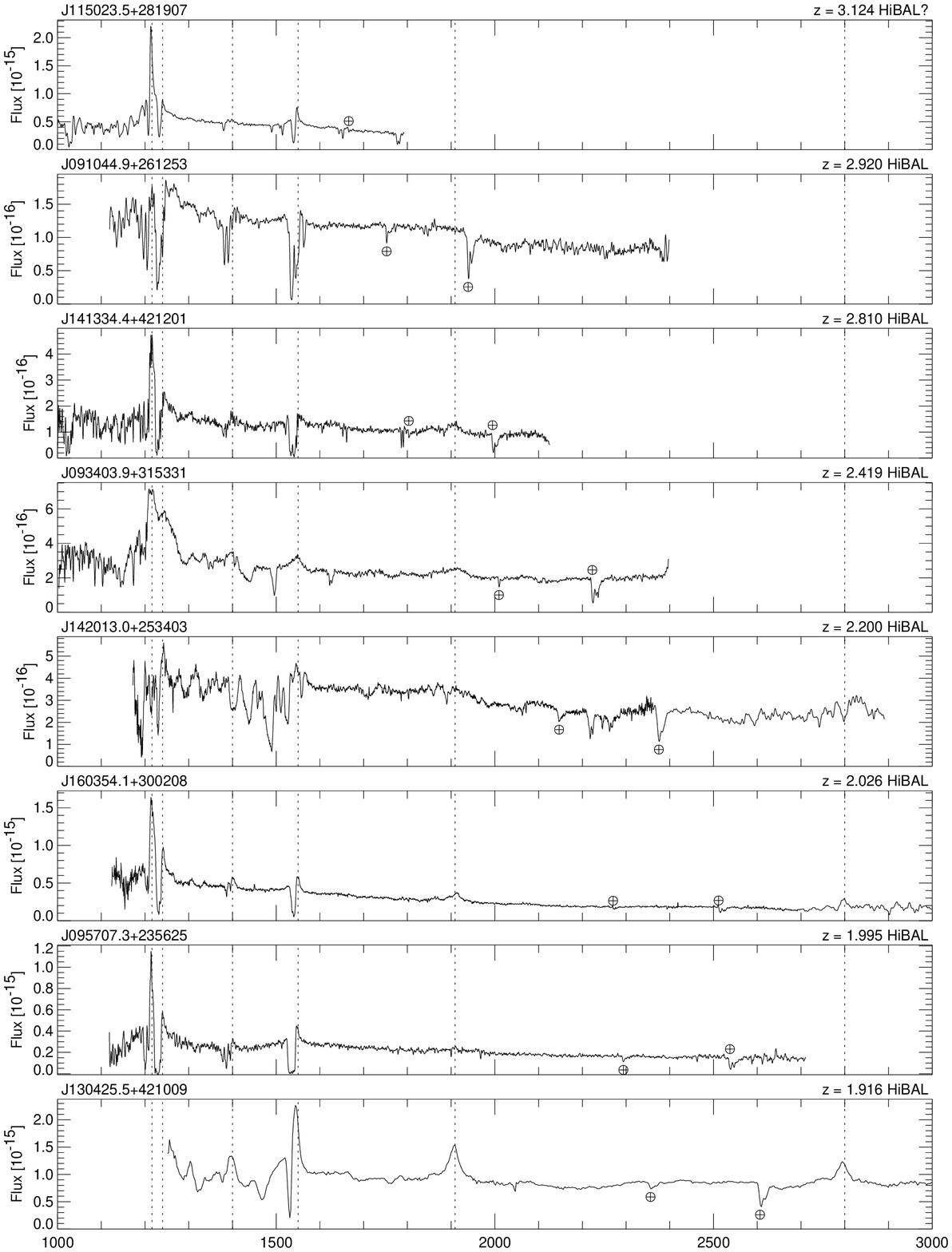}
\caption{Spectra of FBQS high-ionization BAL quasars.
\label{fighispectra}}
\end{figure*}

\begin{figure*}[p]
\figurenum{2}
\epsfxsize=\textwidth \epsfbox{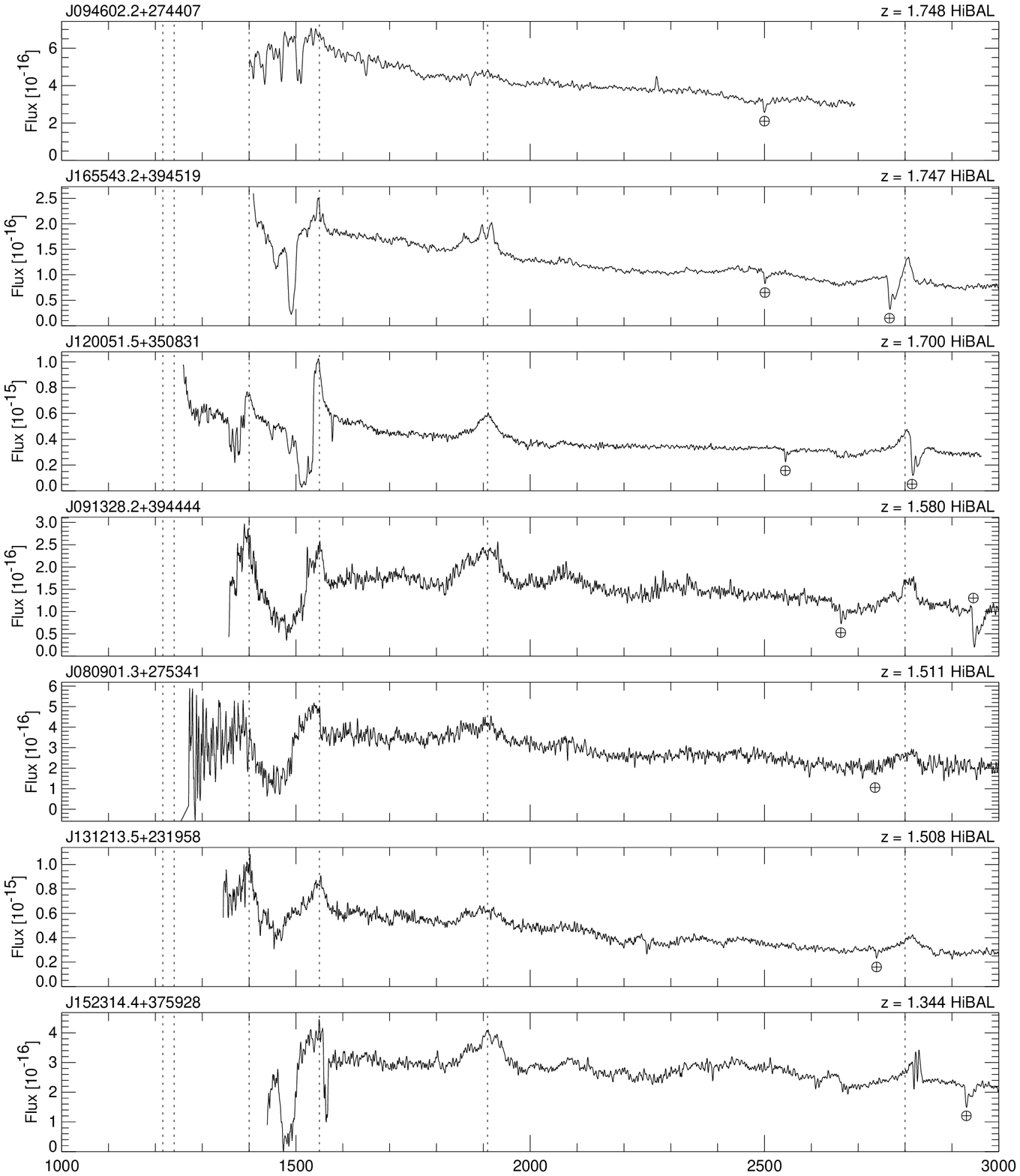}
\caption{{\it Continued.}  Spectra of FBQS high-ionization BAL quasars.}
\end{figure*}
}

The BAL quasars in Table 1 divide nearly evenly into 15 HiBALs and 14
LoBALs.  Both FIRST J105427.2+253600 and J132422.5+245222 are
classified as LoBALs solely by the presence of Al~III $\lambda1860$,
since their Mg~II is redshifted into the near IR.  Four of the LoBALs
belong to the rare class of FeLoBALs (Becker et al.\ 1997),
characterized by the metastable FeII absorption bands centered at
$\sim2350$ and $\sim2575$~\AA.  These four FeLoBAL quasars vary
markedly in the depth of their absorption features.  Two of the other
LoBAL quasars (FIRST~J140806.2+305449 and J152350.4+391405) are
unusual insofar as the spectra appear suppressed blueward of 2500~\AA\
in the rest frame.  Several of the BAL quasars, both high and low
ionization, are almost devoid of obvious emission lines (e.g.,
FIRST~J142703.6+270940, J142013.1+253404).

\section{Radio Properties of the FIRST BAL Quasars}

The FIRST Survey provides 20 cm maps with 5 arcsec angular resolution
taken with the NRAO VLA\footnote{The NRAO is operated by Associated
Universities, Inc., under a cooperative agreement with the National
Science Foundation.} in the B-configuration.  To investigate the
radio spectral indices and radio morphologies of the radio-selected
BAL quasars, we have reobserved nearly all of the objects with the VLA
in either A or D configurations (sometimes both) at 20 and 3.6~cm
wavelength.  The observed flux densities of the quasars at 20~cm from
the three different VLA configurations (A, B, and D) are given in
Table~2, along with the A and D configuration 3.6~cm flux densities.
These are supplemented by data from the WENSS (Westerbork Northern Sky
Survey, \cite{wenss}) survey at 92~cm, the Green~Bank 6~cm survey
(\cite{becker91}), and the NVSS 20~cm survey (Condon et al.\ 1998).
Angular resolutions for all the observations are listed in Table~2.
Spectral indices are given for 28 of the BAL quasars; the sources have
a mix of flat spectra (9 sources, $\alpha > -0.5$) and steep spectra
(19 sources, $\alpha \le -0.5$), with 9 of the sources falling close
to the dividing line ($-0.6 \le \alpha \le -0.4$). Where possible, the
spectral indices are based on simultaneous observations at two
frequencies.

This heterogeneous set of radio observations spans a wide range of
angular resolutions, so the measured flux densities may not be directly
comparable; any spectral index derived from data with different
angular resolutions is uncertain unless the radio source is
effectively a point source at the highest resolution
available.  In that case, the flux densities from all the observations
can be directly compared, assuming a nonvariable source.  If all the
flux density measurements at a given frequency for an object are the
same independent of angular resolution, then the point source
assumption is probably valid.  If the flux density measured with low
resolution is less than that at high-resolution, then the source is
probably variable.  If the low-resolution flux density is higher than
the high-resolution value, the difference could arise from either resolution
effects or variability.

\placetable{table2}


\preprint{

\begingroup
\tabcolsep=3pt
\begin{deluxetable}{lcccccccccl}
\tablewidth{0pt}
\tablecaption{Radio Properties of FIRST BAL Quasars}
\tablenum{2}
\tablehead{
\colhead{} & \multicolumn{4}{c}{VLA S(20 cm)} & \colhead{} &
\multicolumn{2}{c}{VLA S(3.6 cm)} & \colhead{} & \colhead{} & \colhead{}\\
\cline{2-5} \cline{7-8}
\colhead{} & \colhead{} & \colhead{} & \colhead{} & \colhead{} & \colhead{} &
\colhead{} & \colhead{} & \colhead{} & \colhead{Spectral} & \colhead{} \\
\colhead{Name} & \colhead{A} & \colhead{B (FIRST)} & \colhead{D (NVSS)} & \colhead{D} & \colhead{} &
\colhead{A} & \colhead{D} & \colhead{} & \colhead{Index} & \multicolumn{1}{c}{Notes} \\
\colhead{(1)} & \colhead{(2)} & \colhead{(3)} & \colhead{(4)} & \colhead{(5)} & \colhead{} &
\colhead{(6)} & \colhead{(7)} & \colhead{} & \colhead{(8)} & \multicolumn{1}{c}{(9)}
}
\startdata
0724+4159 & \nodata & 7.9 & 12.1\phn & 9.9 && \nodata & 10.6\phn && $+0.0$ &  \nl
0728+4026 & 18.0\phn & 17.0\phn & 17.6\phn & \nodata && 2.7 & \nodata && $-1.1$ & $S_{92}=19$ \nl
0809+2753 & \nodata & 1.7 & $<$ & \nodata && \nodata & \nodata && \nodata &  \nl
0840+3633 & 0.9 & 1.6 & 5.2 & 4.9 && 0.5 & 1.1 && $-0.2$ &  \nl
0910+2612 & \nodata & 7.8 & 7.0 & 9.4 && \nodata & 4.2 && $-0.5$ &  \nl
0913+3944 & 2.3 & 2.1 & $<$ & 1.5 && 0.4 & 0.5 && $-0.6$ &  \nl
0934+3153 & 5.1 & 4.7 & 6.5 & \nodata && 3.5 & 3.6 && $-0.2$ &  \nl
0946+2744 & \nodata & 3.6 & 3.9 & 2.6 && \nodata & $<$0.2\phm{$<$} && $<-1.5$ &  \nl
0957+2356 & \nodata & 140.\phn\phn\phn & 137.\phn\phn\phn & 139.\phn\phn\phn && \nodata & 51.3\phn && $-0.6$ & $S_6=84$ \nl
1031+3953 & 1.7 & 2.5 & 3.0 & 1.6 && 1.2 & 1.2 && $-0.2$ &  \nl
1044+3656 & 15.6\phn & 15.0\phn & 15.8\phn & 14.1\phn && 5.2 & 5.9 && $-0.5$ & $S_{92}=39$ \nl
1054+2536 & 3.0 & 3.0 & 2.9 & 2.6 && 0.9 & 1.1 && $-0.5$ &  \nl
1122+3124 & \nodata & 12.9\phn & 10.9\phn & 10.\phn\phn && \nodata & 3.6 && $-0.6$ & $S_{92}=23$ \nl
1150+2819 & \nodata & 14.2\phn & 11.0\phn & 9.7 && \nodata & 1.3 && $-1.2$ &  \nl
1200+3508 & 1.8 & 2.0 & 2.9 & 1.6 && $<$0.2\phm{$<$} & 0.4 && $-0.8$ &  \nl
1214+2803 & $<$1.8\phm{$<$} & 2.9 & $<$ & 2.1 && $<$0.4\phm{$<$} & 0.5 && $-0.8$ &  \nl
1304+4210 & \nodata & 1.5 & $<$ & 0.9 && \nodata & 2.8 && $+0.7$ &  \nl
1312+2319 & 46.5\phn & 44.1\phn & 46.9\phn & 45.5\phn && 12.6\phn & 12.3\phn && $-0.8$ &  \nl
1324+2452 & \nodata & 4.9 & 4.4 & 4.3 && \nodata & 1.2 && $-0.7$ &  \nl
1408+3451 & 2.7 & 2.9 & 3.1 & 1.7 && 0.4 & 0.6 && $-0.6$ &  \nl
1408+3054 & 3.0 & 3.3 & 4.2 & 1.3 && 0.4 & 0.4 && $-0.7$ &  \nl
1413+4212 & \nodata & 18.7\phn & 17.3\phn & 17.0\phn && \nodata & 11.3\phn && $-0.2$ & $S_{92}=22$ \nl
1420+2534 & 0.5 & 1.3 & $<$ & \nodata && $<$0.2\phm{$<$} & 0.2 && $-1.1$ &  \nl
1427+2709 & $<$3.\phm{$<$} & 3.0 & $<$ & 1.9 && 0.3 & 0.6 && $-0.7$ &  \nl
1523+3759 & 1.3 & 1.8 & $<$ & 0.6 && $<$0.2\phm{$<$} & 0.2 && $-0.6$ &  \nl
1523+3914 & 4.1 & 4.1 & 4.3 & 2.8 && 1.5 & 1.5 && $-0.4$ & $S_{92}=18$ \nl
1603+3002 & 54.2\phn & 54.2\phn & 54.6\phn & 50.7\phn && 18.1\phn & 18.6\phn && $-0.6$ & $S_{92}=43$ \nl
1641+3058 & \nodata & 2.7 & 3.4 & 1.1 && \nodata & 2.5 && $+0.5$ &  \nl
1655+3945 & \nodata & 10.2\phn & 10.5\phn & 4.8 && \nodata & 3.4 && $-0.2$ &  \nl
\enddata
\footnotesize
\tablecomments{All flux densities are in mJy.  Descriptions of table columns:\\
Col 1: Truncated BAL QSO source name.\\
Col 2: 20~cm flux density measured by VLA in A-configuration
(resolution 1.8\arcsec).\\
Col 3: 20~cm flux density measured by VLA in B-configuration
(resolution 5.4\arcsec) from FIRST catalog (White et al.\ 1997).\\
Col 4: 20~cm flux density measured by VLA in D-configuration
(resolution 45\arcsec) from NVSS catalog (Condon et al.\ 1998).  ``$<$'' indicates
source is not detected.\\
Col 5: 20~cm flux density measured by VLA in D-configuration
in November 1997.\\
Col 6: 3.6~cm flux density measured by VLA in A-configuration
(resolution 0.3\arcsec).\\
Col 7: 3.6~cm flux density measured by VLA in D-configuration
(resolution 9\arcsec).\\
Col 8: Spectral index between 3.6 and 20~cm ($F_\nu \propto \nu^\alpha$).\\
Col 9: Notes: $S_{92}$ is 92~cm flux density from WENSS survey (Rengelink et
al.\ 1997).
$S_6$ is 6~cm flux density from Green Bank survey (Becker, White, \& Edwards
1991).
}
\end{deluxetable}
\endgroup


}

The VLA observations provide some indication of the radio brightness
distribution and morphology. Roughly 90\% of the BAL
quasars appear point-like at the FIRST resolution of $\sim5\arcsec$.
This is in sharp contrast to the parent population of quasars in the
FBQS, which are evenly split between point-like and extended (based on
a subset of several hundred quasars with a similar range of parameters to the BAL
quasars:  $z>0.5$ and a 20~cm flux density less than 50~mJy).  Of
13 BAL quasars observed in the A configuration of the VLA, 11 are
still unresolved at the level of $\sim1.5\arcsec$.

\subsection{Comments on Individual Radio Sources}

FIRST~J072418.4+415914 -- Appears to be variable at 20~cm.

FIRST~J080901.3+275342 -- Slightly resolved in FIRST.

FIRST~J084044.5+363328 -- A second radio source positioned $\sim27\arcsec$ away
from 0840+3633 has a FIRST flux density of 2.5 mJy and appears extended. 
This source is too close for the NVSS to resolve from 0840+3633 and
probably explains the higher NVSS flux density. The lower B-configuration 20 cm
flux density was used to determine the spectral index in Table 2.

FIRST~J093404.0+315331 -- Appears to be variable at 20~cm.

FIRST~J115023.6+281907 -- Appears to be variable at 20~cm.

FIRST~J140806.2+305449 -- Possibly a triple source, although a chance alignment of
sources is more likely since both of the other sources break
up into double sources in higher resolution images.

FIRST~J152350.4+391405 -- Appears to be variable at 20~cm.

FIRST~J160354.2+300209 -- Possibly a GigaHertz-peaked radio spectrum.

FIRST~J164152.3+305852 -- Partially resolved by FIRST, consistent with the higher NVSS
flux density but subsequent D~configuration data suggests variability.

FIRST~J165543.2+394520 -- Appears to be variable at 20~cm.

\section{The FBQS BAL Quasar Fraction and Its Dependence on Radio-Loudness}

The frequency of BALs within the FBQS can be derived by comparing the number of
BAL quasars found to the number of quasars with rest-frame wavelength coverage
(determined by the redshift and observed spectral range) that would have
allowed absorption to be seen had it been present.  For the redshift range
relevant to LoBAL quasars, $0.5 \lesssim z \lesssim 1.7$, there are $\sim350$
quasars in the FBQS in which LoBALs could have been confirmed had they been
present.  In this same redshift range, we find 11 are LoBAL or FeLoBAL quasars
(3$\pm$1\%, where the uncertainty is the standard deviation of a binomial
distribution).  There are 100 quasars in the redshift range relevant to HiBALs,
$z \gtrsim 1.4$, in which the wavelength coverage would have permitted C~IV
absorption to be seen had it been present.  Of these 100 quasars, eighteen show
high-ionization broad absorption.  This includes the 3 high-redshift LoBALs, so
designated because they also show Al III absorption; the other 15
objects show only high-ionization absorption.  Our BAL rate is therefore
$18\pm3.8$\%.  If we exclude from this those objects with zero BALnicity,
our rate is reduced to 14\% which is roughly a 50\% increase over the
rate seen in optically selected samples ($\sim$9\% in the LBQS; Foltz et
al.\ 1990).  It is worth noting of the unambiguous BAL quasars, i.e., those
with nonzero BALnicity, one third are either LoBALs or FeLoBALs while the
comparable number for the LBQS is 10\%. This is very suggestive that the
frequency of LoBAL quasars is highly dependent on radio luminosity as was
already postulated in \cite{becker97}.

Until the discovery of FIRST~J155633.8+351758 by Becker et al.\ (1997)
and several additional objects by Brotherton et al.\ (1998), it was
believed that the BAL phenomenon did not occur in radio-loud quasars,
i.e., those with $R^* > 10$.  The 29 FBQS BALs demonstrate otherwise.
In Figure~\ref{figrz} we plot $\log R^*$ vs $z$ (where $R^*$ is taken
from  \cite{white99}).  Consistent with the Becker et al.\ (1997) and
Brotherton et al.\ (1998) results, the BAL quasars are not confined to
the radio-quiet regime.  Our data do suggest that the incidence of BALs
decreases for radio-loud quasars with $R^* > 100$.  For the LoBALs in
particular, $R^*$ never exceeds 35.  In comparison, 38\%  of the
quasars in the FBQS over the same redshift range (z $> 0.5$) have $R^*
> 35$.  The incidence of LoBALs is 5\% for quasars with $R^* < 35$.
While based on a rather small sample, these statistics suggest that the
frequency of LoBAL quasars is dependent on radio loudness. The
likelihood of a quasar being a HiBAL, however, shows no obvious
dependence on radio loudness.  HiBALs are slightly under-represented in
quasars with $R^* > 100$, but this is not statistically significant.  A
better delineation of the frequency of BALs as a function of $R^*$ will
have to wait until more sky is surveyed by the FBQS, but the existence
of a population of radio-loud BALs, both low and high ionization, is
now firmly established.

\placefigure{figrz}

\preprint{
\begin{figure*}
\plotone{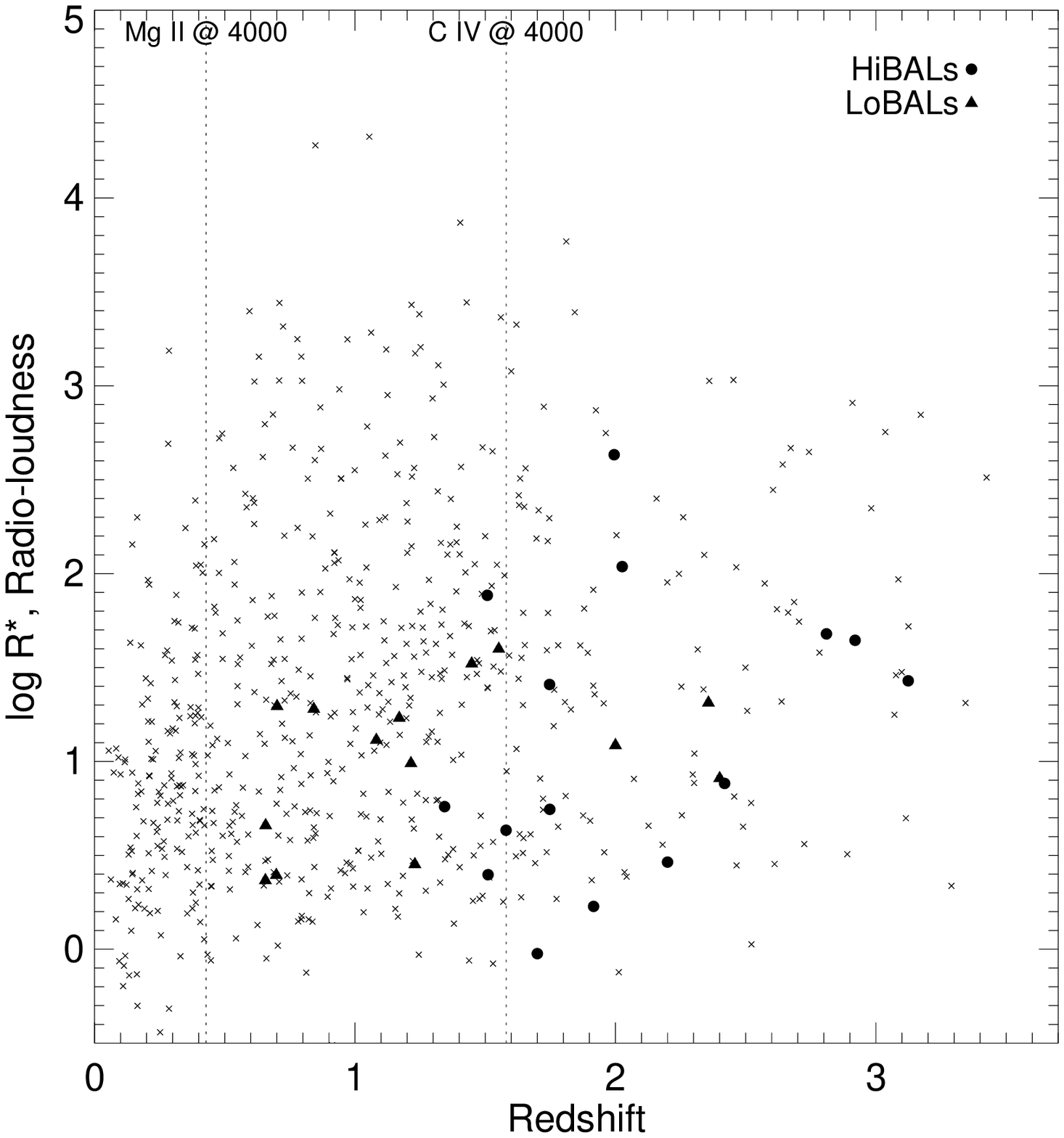}
\caption{ Radio-optical ratio $R^*$ (rest-frame ratio of the 5~GHz
radio flux density to the 2500~\AA\ optical flux; \cite{stocke92})
versus redshift for FBQS quasars.  BAL quasars are indicated with the
heavy symbols, with low-ionization (LoBALs) and high-ionization
(HiBALs) object distinguished.  The dotted lines show redshifts where
Mg~II~$\lambda2800$ and C~IV~$\lambda1549$ fall at 4000~\AA\ in the
observed spectrum, so that their absorption is observable in most of
our spectra (which vary in their wavelength coverage.)
It is difficult or impossible to identify any BAL
systems in quasars with redshifts below $\sim0.4$, and HiBALs
typically cannot be detected when $z\lesssim1.4$.  Many FIRST-selected
BAL quasars are found above the traditional radio-loudness threshold
at $\log R^*=1$.  There also is an evident excess of LoBAL quasars at
radio-intermediate $\log R^* \sim 1$.  }
\label{figrz}
\end{figure*}
}

Using $R^*$ as a measure of radio loudness may be a little misleading for BAL quasars in so
far as the BAL quasars are affected by reddening which would reduce
the optical magnitude and hence inflate the value of $R^*$. As we point out
in section 4.2 (see Figure 6), if we use the alternative definition of
radio-loud, ie, $L_R > 10^{32}$ erg s$^{-1}$ Hz$^{-1}$ (Miller, Rawlings, and
Saunders 1993), which is independent of the observed optical magnitude, we still
find that a significant number of the FBQS BAL quasars are radio-loud.

The traditional measure of the significance of the broad absorption
lines in a quasar spectrum is the BALnicity index (BI;
\cite{weymann91}).  In Figures~\ref{figbalnicity}(a) and
\ref{figbalnicity}(b) we plot the dependence of BI on the observed
radio luminosity, for HiBAL and LoBAL quasars respectively.  For HiBAL
quasars, there is an anticorrelation between BI and $L_R$ (The
Spearman rank correlation coefficient is $-0.85$ which would arise by
chance with a probability of only $6 \times 10^{-5}$).  No such
anticorrelation is apparent for the LoBAL quasars in
Figure~\ref{figbalnicity}(b).  In Figures~\ref{figbalnicity}(c) and
\ref{figbalnicity}(d), we plot the dependence of the maximum outflow
velocity $V_{max}$ against $L_R$ for HiBAL and LoBAL quasars.  There
is a suggestion of an anticorrelation for HiBAL quasars (the Spearman
rank correlation coefficient is $-0.70$; probability of only 0.0037),
though it is considerably less convincing than that with BI.

\placefigure{figbalnicity}

\preprint{
\begin{figure*}
\centering \leavevmode
\epsfxsize=0.7\textwidth \epsfbox{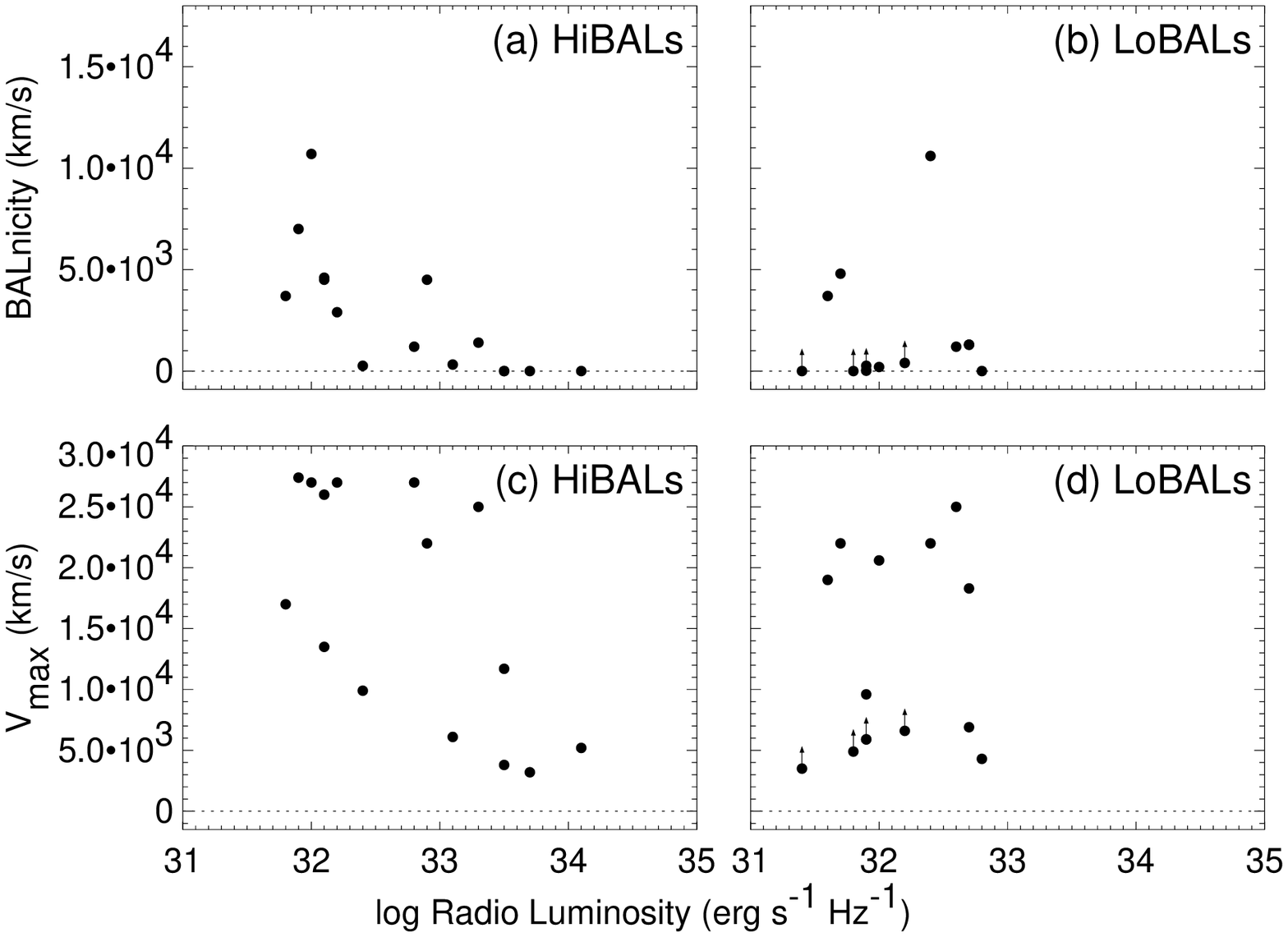}
\caption{
(a) BALnicity index (BI) versus radio luminosity $L_R$ for HiBAL
quasars. (b) BI versus $L_R$ for LoBAL quasars.  (c) Maximum outflow
velocity $V_{max}$ versus $L_R$ for HiBAL quasars.  (d) $V_{max}$
versus $L_R$ for LoBAL quasars.  The FeLoBAL quasars are indicated as
lower limits in (b) and (d) because their complex absorption spectra
make it very difficult to determine these quantities.
}
\label{figbalnicity}
\end{figure*}
}

The lack of correlation between BI or $V_{max}$ and $L_R$ for the
LoBAL quasars may simply reflect the lack of high radio luminosity
LoBAL quasars.  If HiBAL quasars with luminosities greater than
$10^{33}$ ergs/s/Hz are omitted from the plots, at most a weak
correlation is detectable in the less luminous objects.

\subsection{Why the FBQS BAL Quasar Fraction is High}

There are several possible reasons for the higher frequency of BAL
quasars in the FBQS.  One possible explanation is the looser
definition of BAL used in this paper, a definition divorced from the
BALnicity index.  Since the fraction of BALs seen in the LBQS has only
appeared as an AAS abstract (\cite{foltz90}), it is difficult to
evaluate the magnitude of this effect.  Another possible explanation
is that the frequency of the BAL phenomenon depends on the radio
emission, albeit a reversal of the old thesis that only radio-quiet
quasars can have BALs (\cite{stocke92}).  The FBQS quasars span the
radio-quiet/radio-loud boundary, filling in what used to be considered
a bimodal distribution which was perhaps the result of selection
effects in other surveys (White et al.\ 2000).  Based on a limited
sample, Francis, Hooper, \& Impey (1993) found that BAL quasars in the
LBQS appeared primarily in this radio-intermediate regime; accepting
their result as correct, the FBQS naturally includes BALs passed over,
for whatever reasons, by optical surveys.  It is easy to imagine that
the objects in Figure~\ref{figlospectra} that do not have strong
emission lines or that have significantly redder continua than typical
quasars would be overlooked in surveys with optical selection
criteria.

A related reason for the high incidence of BAL quasars is that our
sample was selected using the red $E$ magnitude of the optical
counterparts while most quasar surveys (including the LBQS) are based
on bluer $B$ magnitudes.  A plot of the color of the FBQS quasars as a
function of redshift is shown in Figure~\ref{figcolorz}.  The reddest
objects in the figure are low-redshift objects, in which there is
undoubtedly a large contribution of starlight.  The BALs in general and
the LoBALs in particular are
predominant among the reddest quasars with $z \gtrsim 0.5$, accounting
for over 50\% of the quasars redder than $O-E$ of 1.3.  The BAL
quasars are redder than the average FBQS quasar by $\sim0.5$
magnitude. Hence samples based on $B$\ magnitudes have an effective
magnitude cutoff 0.5 magnitudes higher for BAL quasars than for the
non-BAL quasars which, owing to the steep quasar number counts
would substantially reduce the observed incidence of
BAL quasars. This effect is tantamount to a differential $k$-correction
between BAL and non-BAL quasars and has been discussed in earlier
studies (Boroson \& Meyers 1992; Sprayberry \& Foltz 1992).  It is
possible that the red FBQS BAL quasars represent the tip of the
iceberg and that there remains a large population of BAL quasars that are
yet redder and do not make it into the FBQS despite its weak color
selection ($O-E < 2$) (Becker at al.\ 1997).

\placefigure{figcolorz}

\preprint{
\begin{figure*}
\plotone{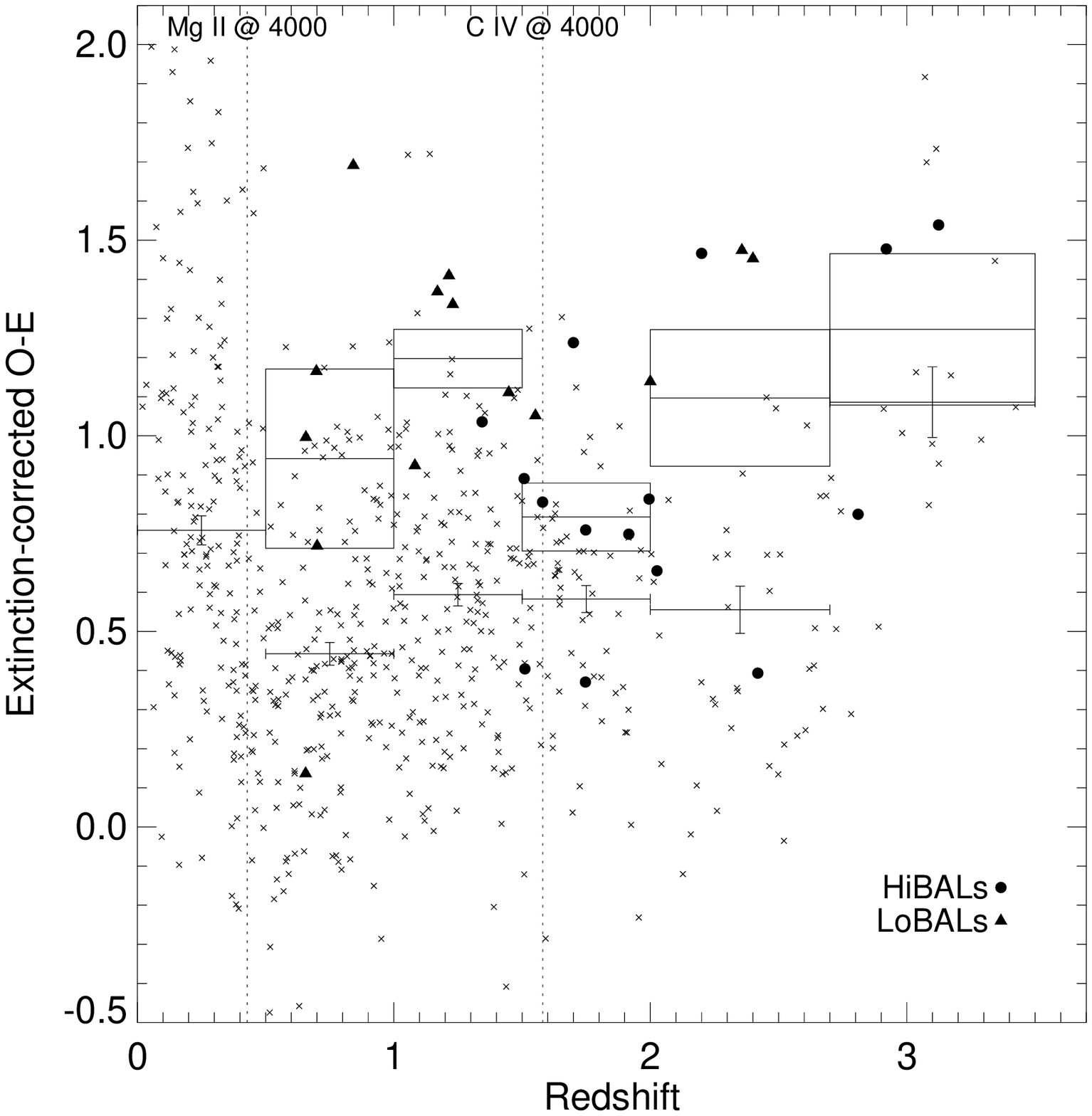}
\caption{ Extinction-corrected $O-E$ (roughly $B-R$) color of FBQS
quasars as a function of redshift.  BAL quasars are marked as in
Fig.~3.  The error bars show the mean and standard deviation of the
mean for the FBQS sample as a whole in redshift bins. The boxes show
the same for the BAL quasars, which are substantially redder.  BAL
quasars constitute the majority of FBQS quasars redder than $O-E=1.3$
with $z\gtrsim0.5$.  BALs cannot be observed in lower $z$ quasars, but
they are also reddened by the starlight visible in lower-luminosity
AGN.  }
\label{figcolorz}
\end{figure*}
}

The red colors of the BAL quasars are not simply the effect of broad
absorption lines suppressing the flux in the $O$\ band (typically
$\sim$0.2 mag), although this contributes part of the difference.  The
unabsorbed continuum itself appears red, especially in the LoBAL
quasars, which is suggestive of dust (Brotherton et al.\ 1999;
Yamamoto \& Vansevicius 1999).  The color difference extends to the
infrared (Hall et al.\ 1997).  Egami et al.\ (1996) presented the
near-IR spectrum of Q 0059$-$2735, which displays a very large Balmer
decrement (H$\alpha$/H$\beta$ = 7.6), almost the same as that seen in
FIRST~J155633.8+351758 (Dey 1998, priv. comm.). For case B recombination, H$\alpha$/H$\beta$ = 2.85;
and ``normal'' blue quasars usually show H$\alpha$/H$\beta$ $<$ 4.
(While case B is not likely to apply to Balmer lines and the Balmer decrement,
the empirical result is that the smallest Balmer decrements are consistent with case B
and that the Balmer decrement has been shown to correlate with the
continuum slope in a manner consistent with an intrinsic case B ratio and
dust reddening (Baker 1997)).
The observed Balmer decrements of these extreme objects then imply
$A_V\sim3$, which is consistent with the Brotherton et al.\ (1997)
estimate for J155633.8+351758 based on spectropolarimetry.  Because of the
bright magnitude limit of the FBQS ($E=17.8$), and the redshifts
required to see BALs in the optical, modest reddening will remove BAL
quasars from the sample (especially the LoBAL quasars which appear
redder than other classes, e.g., Sprayberry and Foltz (1992)).  If it suffered no intrinsic reddening,
FIRST~J155633.8+351758 would likely have been selected for inclusion in the FBQS.



\subsection{Why the FBQS BAL Quasar Fraction is Low}

The FBQS certainly misses BAL quasars.  The magnitude limit
discriminates against BAL quasars when the heavily absorbed spectral
regions fall within the $E$\ bandpass ($\sim 6250 \pm 180$~\AA).
While less affected than the $O$ bandpass, $E$\ magnitudes are still
affected by dust reddening.  The BAL quasar fraction we find,
$\sim$18\%, is then a lower limit to the actual BAL quasar fraction
for radio-intermediate quasars.  The LoBAL quasars, which can be
significantly dust reddened, are more susceptible to
color selection effects than HiBAL quasars.

Goodrich (1997) and Krolik \& Voit (1998) have argued that the true
fraction of BAL quasars is much higher ($\gtrsim$30\%).  While we
would agree that the true fraction is possibly this high, the FBQS
sample undermines their position that BAL quasars are radio-moderate
(\cite{fhi93}) because of optical attenuation rather than intrinsically strong
radio emission.  Figure~\ref{figradio_z} plots the radio luminosity of
the FBQS quasars as a function of redshift, clearly showing that at
least some BAL quasars are intrinsically strong radio sources.

\placefigure{figradio_z}

\preprint{
\begin{figure*}
\plotone{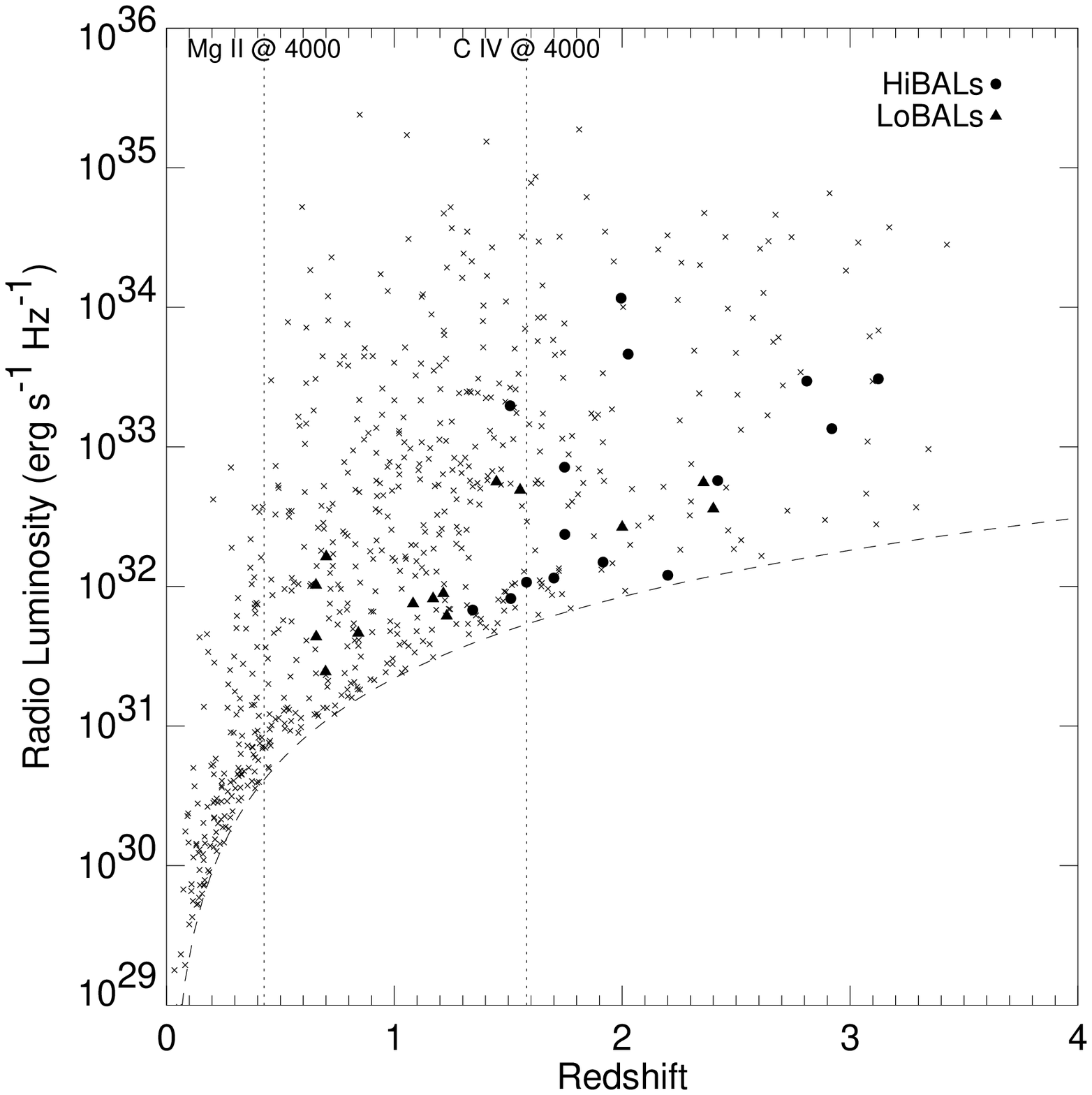}
\caption{
Radio luminosity at a rest frequency of 5~GHz versus redshift,
using spectral index $\alpha=-0.5$ for the non-BAL quasars and
the observed spectral indices from Table~2 for the BAL quasars.
}
\label{figradio_z}
\end{figure*}
}

Goodrich (1997) (see also Goodrich \& Miller 1995) had a second reason
for arguing that the optical continuum was suppressed and the true
fraction of BAL quasars was underestimated: high polarization.  The
idea is that scattered and polarized light is present in all quasars,
but only becomes noticeable when the direct light is somehow
attenuated.  Hutsemekers, Lamy, \& Remy (1998) found that, on average,
LoBAL quasars are more polarized than HiBAL quasars, which are in turn
(slightly) more polarized than non-BAL quasars.  This is consistent
with the idea that LoBAL quasars possess more absorbing material and dust along
the line of sight. 

Estimating the true fraction of BAL quasars remains a difficult
problem given the many unknowns that must be assumed or derived with
incomplete information.  What we can say with some certainty is that
the true fraction of BAL quasars among radio-selected quasars is greater than 18\%.

\subsection{Radio Properties and Unified Schemes}

The similarity of the emission lines in BAL quasars and normal quasars
(\cite{weymann91}) suggests that BAL quasars are normal quasars
seen at a viewing angle that intersects an outflow common to all
quasars.  The spectropolarimetry results have often been interpreted
in terms of a preferred orientation: Goodrich \& Miller (1995), Hines
\& Wills (1995), and Cohen et al.\ (1995) all suggest that BAL quasars are
normal quasars seen along a line of sight skimming the edge of a disk
or torus, with BAL clouds accelerated from its surface by a wind, and
polarized continuum light scattered above along a less obscured path.
LoBAL quasars are those seen at the largest inclinations, thus
presenting the largest column densities.

The jets of quasars provide a way to measure orientation.
The relativistic beaming model for radio sources (e.g., Orr \& Browne
1982) unifies core-dominated (flat spectrum) and lobe-dominated (steep
spectrum) radio sources by means of orientation: core-dominant objects
are those viewed close to the jet axis, while lobe-dominant objects
are those viewed at larger angles. Indeed, relativistic jets appear to be
present in at least some radio-quiet quasars (e.g., Blundell \&
Beasley 1998), and a flat radio spectral index in a radio-quiet quasar
may indicate a beamed source (e.g., Falcke, Sherwood, \& Patnaik
1996).

We find that about two thirds of the FBQS BAL quasars have steep
radio spectra ($\alpha < -0.5$), as expected for edge-on systems, but
that the remaining third have flat spectra (including clearly
radio-loud sources such as FIRST~J141334.4+421202, as well as
J155633.8+351758 which is not in this
sample). This is inconsistent with the simple unified scheme, which predicts
only steep spectrum sources for an edge-on geometry.

Similarly, Barvainis \& Lonsdale (1997) found that the radio spectra
of radio-quiet BAL quasars have a range of slopes, again including
both flat and steep spectra, suggesting that BAL quasars are seen for
a range of orientations with respect to the system (jet) axis.

The radio morphology of the FIRST BAL quasars is also unexpected for
the unified edge-on scheme.  VLA A array maps of our BALQSOs show that
80\% of the sources are unresolved at the 0.2\arcsec\
scale.  This could be because they are small, ``frustrated'' or young
sources, similar to compact steep spectrum sources, or because they
are very core-dominated with the jet beamed toward us.  The
compactness of the radio emission, even in the radio-loud sources,
favors their existence in gas-rich interacting systems which can
confine the radio emission to small scales.

There is an alternative to ``unification by orientation,'' which may
be described as ``unification by time,'' with BAL quasars
characterized as young or recently refueled quasars.  Boroson \&
Meyers (1992) found that LoBAL quasars constitute 10\% of
IR-selected quasars, greater than the 1\% found in
optically selected samples, and that LoBALs show very weak narrow [O III]
$\lambda$5007 emission.  Turnshek et al.\ (1997) found that 1/3 of weak
[O III] $\lambda$5007 quasars show BALs.  Because [O III]
$\lambda$5007 is emitted from the extended narrow-line region (NLR),
its weakness suggests that obscuring material with a large covering
factor is present.  We are unaware of {\em any} LoBAL quasars with
significant [O III] $\lambda$5007 emission.  Voit et al.\ (1993) argue
that low-ionization BALs are a manifestation of a ``quasar's efforts
to expel a thick shroud of gas and dust,'' consistent with the
scenario of Sanders et al.\ (1988) in which quasars emerge from
dusty, gas-rich merger-produced ultraluminous infrared galaxies.
The warm $IRAS$-selected BAL quasars ($0.25 < F_\nu(25\mu
m)/F_\nu(60\mu m)$ $< 3$) Markarian 231 (Smith et al.\ 1995), $IRAS$
07598+6508 (Boyce et al.\ 1996), and PG 1700+518 (Hines et al.\ 1999;
Stockton et al.\ 1998) all show evidence for recent mergers or
interactions, including young starbursts.

While the geometry of BAL quasars and their relationship to non-BAL
quasars remains an open question, our results do not appear to favor
the popular notion that all BAL quasars are normal quasars seen
edge-on.  The FBQS results are more consistent with the
unification by time picture.

\section{Summary}

We have investigated the properties of 29 radio-selected BAL quasars
found in the FBQS.  The sample comprises 15 high-ionization BAL
quasars, and 14 low-ionization BAL quasars, 4 of which are rare
FeLoBALs.  At least 13 are formally radio-loud, unequivocally
establishing the existence of a substantial population of radio-loud
quasars exhibiting BAL spectral features.

The frequency of BAL quasars appears to be higher than that found in optically
selected samples.  Even so, because of selection effects and
preferential reddening of LoBAL quasars, the FBQS almost certainly
misses additional BAL quasars and the true frequency must be higher.
The situation is complicated by indications that the frequency of BAL
quasars peaks among the radio-moderate population and decreases for
the extremes of radio-loudness.  The BAL quasars show compact radio
morphologies, and have a range in radio spectral indices.  The radio
properties do not support the popular scenario in which all BAL
quasars are normal quasars seen edge-on.  An alternative picture in
which BALs are an early stage in the development of new or refueled
quasars is preferred.

\acknowledgments

The success of the FIRST survey is in large measure due to the generous
support of a number of organizations.  In particular, we acknowledge
support from the NRAO, the NSF (grants AST-98-02791 and AST-98-02732),
the Institute of Geophysics and Planetary Physics (operated under the
auspices of the U.S. Department of Energy by the University of California
Lawrence Livermore National Laboratory under contract No.~W-7405-Eng-48),
the Space Telescope Science Institute, NATO, the National Geographic Society
(grant NGS No.~5393-094), Columbia University, and Sun Microsystems.
We also acknowledge several very helpful comments from an anonymous referee.

\preprint{\end{document}}

\newpage

\figcaption[fig1a.eps,fig1b.eps]{ Spectra of low ionization broad absorption line quasars
from the FIRST Bright Quasar Survey, sorted by decreasing redshift, plotted
against rest wavelength.  The dotted lines show expected positions of
prominent emission lines: Ly$\alpha$~1216, N~V~1240, Si~IV~1400,
C~IV~1549, C~III]~1909, and Mg~II~2800.  The positions of the
atmospheric A and B band absorption at observed wavelengths of
$\sim6880$~\AA\ and 7620~\AA\ are marked.
\label{figlospectra}}

\figcaption[fig2a.eps,fig2b.eps]{Spectra of FBQS high-ionization BAL quasars.
\label{fighispectra}}

\figcaption[fig3.eps]{
Radio-optical ratio $R^*$ (rest-frame ratio of the 5~GHz
radio flux density to the 2500~\AA\ optical flux; \cite{stocke92})
versus redshift for FBQS quasars.  BAL quasars are indicated with the
heavy symbols, with low-ionization (LoBALs) and high-ionization
(HiBALs) object distinguished.  The dotted lines show redshifts where
Mg~II~$\lambda2800$ and C~IV~$\lambda1549$ fall at 4000~\AA\ in the
observed spectrum, so that their absorption is observable in most of
our spectra (which vary in their wavelength coverage.)
It is difficult or impossible to identify any BAL
systems in quasars with redshifts below $\sim0.4$, and HiBALs
typically cannot be detected when $z\lesssim1.4$.  Many FIRST-selected
BAL quasars are found above the traditional radio-loudness threshold
at $\log R^*=1$.  There also is an evident excess of LoBAL quasars at
radio-intermediate $\log R^* \sim 1$.
\label{figrz}}

\figcaption[fig4.eps]{
(a) BALnicity index (BI) versus radio luminosity $L_R$ for HiBAL
quasars. (b) BI versus $L_R$ for LoBAL quasars.  (c) Maximum outflow
velocity $V_{max}$ versus $L_R$ for HiBAL quasars.  (d) $V_{max}$
versus $L_R$ for LoBAL quasars.  The FeLoBAL quasars are indicated as
lower limits in (b) and (d) because their complex absorption spectra
make it very difficult to determine these quantities.
\label{figbalnicity}}

\figcaption[fig5.eps]{ Extinction-corrected $O-E$ (roughly $B-R$) color of FBQS
quasars as a function of redshift.  BAL quasars are marked as in
Fig.~3.  The error bars show the mean and standard deviation of the
mean for the FBQS sample as a whole in redshift bins. The boxes show
the same for the BAL quasars, which are substantially redder.  BAL
quasars constitute the majority of FBQS quasars redder than $O-E=1.3$
with $z\gtrsim0.5$.  BALs cannot be observed in lower $z$ quasars, but
they are also reddened by the starlight visible in lower-luminosity
AGN.\label{figcolorz}}

\figcaption[fig6.eps]{ Radio luminosity at a rest frequency of 5~GHz versus redshift,
using spectral index $\alpha=-0.5$ for the non-BAL quasars and the observed
spectral indices from Table~2 for the BAL quasars.\label{figradio_z}}

\end{document}